\documentclass[12pt]{article}

\def\hybrid{\topmargin -20pt    \oddsidemargin 0pt
        \headheight 0pt \headsep 0pt
        \textwidth 6.25in       
        \textheight 9.5in       
        \marginparwidth .875in
        \parskip 5pt plus 1pt   \jot = 1.5ex}

\hybrid

\catcode`\@=11

\def\marginnote#1{}
\newcount\hour
\newcount\minute
\newtoks\amorpm
\hour=\time\divide\hour by60
\minute=\time{\multiply\hour by60 \global\advance\minute by-\hour}
\edef\standardtime{{\ifnum\hour<12 \global\amorpm={am}%
        \else\global\amorpm={pm}\advance\hour by-12 \fi
        \ifnum\hour=0 \hour=12 \fi
        \number\hour:\ifnum\minute<10 0\fi\number\minute\the\amorpm}}
\edef\militarytime{\number\hour:\ifnum\minute<10 0\fi\number\minute}

\def\draftlabel#1{{\@bsphack\if@filesw {\let\thepage\relax
   \xdef\@gtempa{\write\@auxout{\string
      \newlabel{#1}{{\@currentlabel}{\thepage}}}}}\@gtempa
   \if@nobreak \ifvmode\nobreak\fi\fi\fi\@esphack}
        \gdef\@eqnlabel{#1}}
\def\@eqnlabel{}
\def\@vacuum{}
\def\draftmarginnote#1{\marginpar{\raggedright\scriptsize\tt#1}}

\def\draft{\oddsidemargin -.5truein
        \def\@oddfoot{\sl preliminary draft \hfil
        \rm\thepage\hfil\sl\today\quad\militarytime}
        \let\@evenfoot\@oddfoot \overfullrule 3pt
        \let\label=\draftlabel
        \let\marginnote=\draftmarginnote
   \def\@eqnnum{(\theequation)\rlap{\kern\marginparsep\tt\@eqnlabel}%
\global\let\@eqnlabel\@vacuum}  }

\def\preprint{\twocolumn\sloppy\flushbottom\parindent 2em
        \leftmargini 2em\leftmarginv .5em\leftmarginvi .5em
        \oddsidemargin -.5in    \evensidemargin -.5in
        \columnsep .4in \footheight 0pt
        \textwidth 10.in        \topmargin  -.4in
        \headheight 12pt \topskip .4in
        \textheight 6.9in \footskip 0pt
        \def\@oddhead{\thepage\hfil\addtocounter{page}{1}\thepage}
        \let\@evenhead\@oddhead \def\@oddfoot{} \def\@evenfoot{} }

\def\numberbysection{\@addtoreset{equation}{section}
        \def\theequation{\thesection.\arabic{equation}}}

\def\underline#1{\relax\ifmmode\@@underline#1\else
        $\@@underline{\hbox{#1}}$\relax\fi}

\def\titlepage{\@restonecolfalse\if@twocolumn\@restonecoltrue\onecolumn
     \else \newpage \fi \thispagestyle{empty}\c@page\z@
        \def\thefootnote{\fnsymbol{footnote}} }

\def\endtitlepage{\if@restonecol\twocolumn \else \newpage \fi
        \def\thefootnote{\arabic{footnote}}
        \setcounter{footnote}{0}}  

\catcode`@=12
\relax

\makeatletter
\newcounter{pubctr}
\def\publist{\@ifnextchar[{\@publist}{\@@publist}}
\def\@publist[#1]{\list
        {[\arabic{pubctr}]\hfill}{\settowidth\labelwidth{[999]}
        \leftmargin\labelwidth
        \advance\leftmargin\labelsep
        \@nmbrlisttrue\def\@listctr{pubctr}
        \setcounter{pubctr}{#1}\addtocounter{pubctr}{-1}}}
\def\@@publist{\list
        {[\arabic{pubctr}]\hfill}{\settowidth\labelwidth{[999]}
        \leftmargin\labelwidth
        \advance\leftmargin\labelsep
        \@nmbrlisttrue\def\@listctr{pubctr}}}
 \relax
\makeatother
\newskip\humongous \humongous=0pt plus 1000pt minus 1000pt

\newif\ifdtup

\relax

\def\d{\partial}

\def\sqr#1#2{{\vcenter{\vbox{\hrule height.#2pt\hbox{\vrule width.#2pt 
height#1pt \kern#1pt \vrule width.#2pt}\hrule height.#2pt}}}}

\def\=d{\,{\buildrel\rm def\over =}\,}

\def\i3p{\p32\int d^3p}

\def\As{A\hbox to 1pt{\hss /}}
\def\np4{\int d^4p_1\cdots d^4p_{n-1}\, }

\def\Tr{{\rm Tr}\, }

\def\nx4{\int d^4x_1\ldots d^4x_n\, }

\def\kon#1#2{\vbox{\halign{##&&##\cr
\lower4pt\hbox{$\scriptscriptstyle\vert$}\hrulefill &
\hrulefill\lower4pt\hbox{$\scriptscriptstyle\vert$}\cr $#1$&
$#2$\cr}}}

\def\konv#1#2#3{\hbox{\vrule height12pt depth-1pt}
\vbox{\hrule height12pt width#1cm depth-11.6pt}
\hbox{\vrule height6.5pt depth-0.5pt}
\vbox{\hrule height11pt width#2cm depth-10.6pt\kern5pt
      \hrule height6.5pt width#2cm depth-6.1pt}
\hbox{\vrule height12pt depth-1pt}
\vbox{\hrule height6.5pt width#3cm depth-6.1pt}
\hbox{\vrule height6.5pt depth-0.5pt}}
\def\konu#1#2#3{\hbox{\vrule height12pt depth-1pt}
\vbox{\hrule height1pt width#1cm depth-0.6pt}
\hbox{\vrule height12pt depth-6.5pt}
\vbox{\hrule height6pt width#2cm depth-5.6pt\kern5pt
      \hrule height1pt width#2cm depth-0.6pt}
\hbox{\vrule height12pt depth-6.5pt}
\vbox{\hrule height1pt width#3cm depth-0.6pt}
\hbox{\vrule height12pt depth-1pt}}

\def\konw#1#2#3{\hbox{\vrule height12pt depth-1pt}
\vbox{\hrule height12pt width#1cm depth-11.6pt}
\hbox{\vrule height6.5pt depth-0.5pt}
\vbox{\hrule height12pt width#2cm depth-11.6pt \kern5pt
      \hrule height6.5pt width#2cm depth-6.1pt}
\hbox{\vrule height6.5pt depth-0.5pt}
\vbox{\hrule height12pt width#3cm depth-11.6pt}
\hbox{\vrule height12pt depth-1pt}}

\def\i{{\rm int}}

\def\e{{\rm ext}}

\def\r{{\rm ret}}
\def\a{{\rm av}}

\def\m3{{\mu_1\mu_2\mu_3}}

\def\p{{(+)}}

\def\be{\begin{equation}}       \def\eq{\begin{equation}}
\def\ee{\end{equation}}         \def\eqe{\end{equation}}

\def\bea{\begin{eqnarray}}      \def\eqa{\begin{eqnarray}}
\def\ena{\end{eqnarray}}        \def\eea{\end{eqnarray}}
                                \def\eqae{\end{eqnarray}}

\def\ba{\begin{array}}
\def\ea{\end{array}}
\def\unit{1 \hskip-.3em \raise2pt\hbox{$ \scriptstyle |$ } }

\def\a{\alpha}

\def\d{\delta}
\def\e{\epsilon}           
\def\f{\phi}               

\def\i{\iota}

\def\l{\lambda}
\def\m{\mu}

\def\o{\omega}  
\def\p{\pi}                
\def\r{\rho}                                     
\def\s{\sigma}                                   
\def\t{\tau}

\def\L{\Lambda}
\def\O{\Omega}

\def\cf{{\cal F}}

\def\cq{{\cal Q}}

\def\half{{1 \over 2}}

\def\bop#1{\setbox0=\hbox{$#1M$}\mkern1.5mu
        \vbox{\hrule height0pt depth.04\ht0
        \hbox{\vrule width.04\ht0 height.9\ht0 \kern.9\ht0
        \vrule width.04\ht0}\hrule height.04\ht0}\mkern1.5mu}
\def\pa{\partial}                              

\def\>{\rangle} 

\def\<{\langle} 
\def\Dsl{D \hskip-.6em \raise1pt\hbox{$ / $ } }

\def\sl#1{\rlap{\hbox{$\mskip 1 mu /$}}#1}
\def\leftrightarrowfill{$\mathsurround=0pt \mathord\leftarrow \mkern-6mu
       \cleaders\hbox{$\mkern-2mu \mathord- \mkern-2mu$}\hfill
       \mkern-6mu \mathord\rightarrow$}
\def\dvec#1{\vbox{\ialign{##\crcr
       \leftrightarrowfill\crcr\noalign{\kern-1pt\nointerlineskip}
       $\hfil\displaystyle{#1}\hfil$\crcr}}}          
\def\hook#1{{\vrule height#1pt width0.4pt depth0pt}}
\def\leftrighthookfill#1{$\mathsurround=0pt \mathord\hook#1
       \hrulefill\mathord\hook#1$}
\def\underhook#1{\vtop{\ialign{##\crcr                 
       $\hfil\displaystyle{#1}\hfil$\crcr
       \noalign{\kern-1pt\nointerlineskip\vskip2pt}
       \leftrighthookfill5\crcr}}}
\def\smallunderhook#1{\vtop{\ialign{##\crcr      
       $\hfil\scriptstyle{#1}\hfil$\crcr
       \noalign{\kern-1pt\nointerlineskip\vskip2pt}
       \leftrighthookfill3\crcr}}}

\def\sfrac#1#2{{\vphantom1\smash{\lower.5ex\hbox{\small$#1$}}\over
       \vphantom1\smash{\raise.4ex\hbox{\small$#2$}}}} 
\def\bfrac#1#2{{\vphantom1\smash{\lower.5ex\hbox{$#1$}}\over
       \vphantom1\smash{\raise.3ex\hbox{$#2$}}}}      
\def\afrac#1#2{{\vphantom1\smash{\lower.5ex\hbox{$#1$}}\over#2}}  
\def\on#1#2{{\buildrel{\mkern2.5mu#1\mkern-2.5mu}\over{#2}}}
\def\ddt#1{\on{\hbox{\LARGE .\kern-2pt.}}#1}             
\def\tdt#1{\on{\hbox{\LARGE .\kern-2pt.\kern-2pt.}}#1}   

\def\boxes#1{
       \newcount\num
       \num=1
       \newdimen\downsy
       \downsy=-1.5ex
       \mskip-2.8mu
       \bo
       \loop
       \ifnum\num<#1
       \llap{\raise\num\downsy\hbox{$\bo$}}
       \advance\num by1
       \repeat}
\def\boxup#1#2{\newcount\numup
       \numup=#1
       \advance\numup by-1
       \newdimen\upsy
       \upsy=.75ex
       \mskip2.8mu
       \raise\numup\upsy\hbox{$#2$}}

\newskip\humongous \humongous=0pt plus 1000pt minus 1000pt

\newif\ifdtup

\def\NPB#1#2#3{{\it Nucl. Phys.} {\bf B#1} (#2) #3}

\def\to{\rightarrow}

\def\1ov4{{1\over 4}}

\def\Tr{{\rm Tr}}

\def\pa{\partial}
\def\xx{\times}

\def\ddt{\dot{\t}}

\def\pa{\partial}

\def\xx{\times}

\def\nonu{\nonumber \\{}}
\def\half{{1 \over 2}}

\def\IR{\relax{\rm  I\kern-.18em R}}
\def\inv{^{\raise.15ex\hbox{${\scriptscriptstyle -}$}\kern-.05em 1}}

\begin{document}

\thispagestyle{empty} \setcounter{footnote}{0}
\begin{flushright}
CERN-TH/97-330\\
\vskip -.1 cm
KUL-TF-97/33\\ 
\vskip -.1 cm
hep-th/9711138\\
\end{flushright}

\vskip 1,6cm
\begin{center}
{\large \bf Microscopic derivation of the Bekenstein--Hawking
entropy formula for non-extremal black holes}

  \vskip 2.0cm 
{\bf Konstadinos Sfetsos}
\vskip 0,2cm
{\it Theory Division, CERN \\
CH-1211 Geneva 23, Switzerland\\
{\tt sfetsos@mail.cern.ch}}
  \\
  \vskip .9cm
  {\bf Kostas Skenderis}
  \vskip 0,2cm
  {\it Instituut voor Theoretische Fysica, KU Leuven\\
    Celestijnenlaan 200D, B-3001 Leuven, Belgium\\
    {\tt kostas.skenderis@fys.kuleuven.ac.be}}

\end{center}
\vskip 1 cm 
\centerline {\bf Abstract} 
\vskip 0,2cm
\noindent
We derive the Bekenstein--Hawking entropy formula for 
four- and five-dimensional
non-supersymmetric black holes (which include the Schwarzchild ones)
by counting microscopic states. 
This is achieved by first showing that these black holes are 
U-dual to the three-dimensional black hole
of Banados--Teitelboim--Zanelli and then counting microscopic states 
of the latter following Carlip's approach.
Black holes higher than five-dimensional are also considered.
We discuss the connection of our approach to the D-brane picture.

\vfill

\begin{flushleft}
CERN-TH/97-330 \\
\vskip -.1 cm
KUL-TF-97/33\\
\vskip -.1 cm
November 1997
\end{flushleft}

\newpage

\section{Introduction}

Black holes are one of the most fascinating objects 
in general relativity. Their existence has profound 
implications for gravity in both the classical and the quantum regime. 
Black hole quantum mechanics provides a window into strong coupling 
quantum physics by raising a set of puzzles and questions that any consistent 
quantum theory of gravity should solve. The discovery that the black-hole 
laws are thermodynamical in nature \cite{Beke} implies that there should
be an underlying statistical description of them in terms 
of some microscopic states. In addition, black holes 
can evaporate \cite{Hawk}, which leads to the 
``information loss paradox''. These questions, to a large extent,
remained unanswered for more than twenty years. 

String theory claims to provide a consistent theory of gravity.
One would therefore expect that string theory
provide answers to these questions. 
The strong coupling nature of black-hole physics, however, 
requires an understanding of non-perturbative string theory
that was not available until recently. The situation has changed 
dramatically during the last few years. The duality symmetries 
have led to a new unified picture and provided
a handle into strong coupling physics \cite{stringdual}. 
The discovery of D-branes \cite{dbrhor,JPol} 
has led to remarkable progress in the understanding of the physics of 
extremal black holes. In particular, 
it led to identification and counting of microstates 
for this subset of  black holes \cite{vastro,malda}. 
The result was in exact agreement with the
Bekenstein--Hawking entropy formula. The idea behind these 
computations was to construct a D-brane configuration
with the same quantum numbers as the corresponding black hole
we are interested in. The counting of states is then performed
at weak coupling, where the D-brane description is valid.
The BPS property of these configurations implies that the 
number of states remains unchanged as the string coupling grows.
One, therefore, can extrapolate these results to the black-hole phase. 
In this way states were counted for extremal $4d$ and $5d$ black holes.
Near-extremal black holes were also studied \cite{malda,nnon}. 
The absence of supersymmetry, however, makes these results less rigorous.
For the same reason (i.e. absence of supersymmetry)
the physically most interesting case, namely the case of non-extremal
black holes, is untractable in this framework. Let us mention, however,
that a natural extension of these ideas, as formulated in the
correspondence principle of Polchinski and Horowitz \cite{hor3} (for earlier 
ideas see \cite{susss}) does yield the correct dependence of the entropy 
on the mass and the charges, even if 
it does not provide the numerical coefficient.
Recently, similar results for non-extremal black 
holes were obtained in \cite{matrix} using the M(atrix) 
formulation \cite{BFSS} of M-theory. 

Another important development in the understanding of the statistical origin
of the black-hole entropy (that actually preceded the D-brane 
developments) was Carlip's derivation \cite{carlip} of the
Bekenstein--Hawking  
entropy formula for the three-dimensional black hole
of Banados--Teitelboim--Zanelli (BTZ)\cite{BTZ}. 
The latter solves Einstein's equations in the presence of a negative 
cosmological constant and is, therefore, asymptotically anti-de Sitter.
Soon after its discovery it was shown that the BTZ black hole is actually an 
exact solution of string theory \cite{hor2,kaloper}, namely that there is an 
exact conformal field theory (CFT) associated to it.
Physics in three dimensions is significantly simpler 
than in higher dimensions. In particular, three-dimensional gravity 
can be recast as a Chern--Simons theory \cite{town1, wit}. If the space has
a boundary then the Chern--Simons theory induces a WZW action in this
boundary. The latter describes would-be degrees of freedom that become
dynamical because certain gauge transformations become inadmissible 
due to boundary conditions. Carlip has shown that these degrees of freedom
correctly account for the Bekenstein--Hawking entropy of 
the BTZ black hole.\footnote{The idea that only physical degrees of freedom 
defined in a ``stretched'' horizon may 
account for the black hole entropy has been advocated in \cite{stret}.
In a string theory context it was put forward by A. Sen \cite{Sen},
in order to reconcile the Bekenstein--Hawking entropy 
for extremal electric black holes, 
with the entropy of elementary superstring excitations.}
It is important to realize that 
this result is valid both at extremality and away 
from it. However, the method used seems very particular to three
dimensions (see, however, \cite{ashtekar}). Notice also that all D-brane
results are for black holes of dimension higher than three.
The main reason for this is that in constructing a solution
out of D-branes one usually restricts oneself to at least three overall 
transverse directions, and three-dimensional space-time has two transverse 
directions. If the overall transverse directions are 
less than three, the harmonic functions appearing in the 
D-brane configuration are not bounded at infinity. 

To summarize:the D-branes techniques can be used to derive the 
Bekenstein--Hawking entropy for $4d$ and $5d$ 
supersymmetric black holes, whereas Carlip's approach is not restricted
to supersymmetric black holes, but it seems to apply only to $3d$ ones.
We shall show in this article that one can use the latter to 
study non-extremal $4d$ and $5d$ black holes and, in particular, we will 
derive the Bekenstein--Hawking entropy formula associated to them with the 
correct numerical coefficient. Our considerations also apply 
to higher-dimensional black holes, 
although we have no derivation of the Bekenstein--Hawking entropy
formula in these cases.

Over the last few years a new unifying picture of all five string 
theories and eleven-dimensional supergravity has emerged \cite{stringdual}. 
A central 
r\^{o}le in these developments has been played by the various duality 
symmetries. It is now believed that there exist an underlying 
master theory, the M-theory, that has all string theories 
and eleven-dimensional supergravity \cite{11sugra} as special limits. 
The dualities symmetries can be viewed as some kind of gauge 
symmetry of this theory. Physical quantities should be ``gauge-invariant'',
i.e. U-duality-invariant. Choosing one configuration
among all its U-duals to describe a physical system 
corresponds to choosing a particular ``gauge''. As in usual gauge theories,
some gauges are preferable for answering certain questions than   
others. We shall show below that the $4d$ and $5d$ black 
holes are U-dual to the BTZ black hole (for related 
work, see \cite{hyun}). One may, therefore,
choose the ``BTZ gauge'' in order to answer certain physical questions. 
In particular, we shall
address in detail the question of the statistical origin 
of the entropy. 

The BTZ black hole (for $J \neq 0$) 
is non-singular. One may, therefore, argue that the singularities
in the $4d$ and $5d$ black holes are ``gauge'' artefacts.
In addition, the fact that the BTZ black hole is asymptotically 
anti-de Sitter and the $4d$ and $5d$ black holes are asymptotically flat 
implies that the cosmological constant is a ``gauge-dependent'' notion.
Furthermore, the simplicity of the ``BTZ gauge'' may make tractable 
the study of the final state of black holes.
 
One may wonder at this point how is it possible to connect objects 
of different dimensionality using the U-duality group.
Consider, for concreteness, type-II string theory on a torus.
Then, the U-dual group is considered to be the (discretized) version
of the global symmetry group of the various maximal supergravity 
theories obtained from eleven-dimensional supergravity by toroidal 
compactification in dimensions $d \leq 10$. Therefore, almost by 
definition, the dualities do not change the number of non-compact 
dimensions. For static backgrounds, however, one has, in addition 
to the isometries corresponding to toroidal directions, an extra time-like 
non-compact isometry. This leads to a larger group. 
Consider, for instance, the case of $d$ compact directions. The 
T-duality group is $O(d, d)$. Suppose for a moment that the time is compact
with radius $R$. Then the symmetry group would be enlarged to $O(d+1, d+1)$. 
To see what happens in the decompactification
we let $R$ become larger and larger while restricting the elements 
of $O(d+1, d+1)$ to the ones that do not mix the coordinates with 
finite radii with the time coordinate. In the limit $R \to \infty$ the time 
becomes non-compact and we are left with a subgroup of $O(d+1, d+1)$.
The latter is basically a combination of diffeomorphisms 
of the time coordinate, which involve the compact coordinates
and the $O(d,d)$ transformations of the compact coordinates themselves. 
In particular, this group
contains elements that correspond to isometries that are 
space-like everywhere except at spatial infinity, 
where they become null.
T-dualizing with respect to 
these isometries changes the asymptotic geometry of space-time
\cite{hor2,BPS}.
There seems to be a widespread belief that string theory admits 
only Ricci-flat compactifications. This is, however, not true.
We shall exhibit below exact string solutions that correspond
to compactifications on $S^2$ and $S^3$ times some torus.
T-dualities along the above-mentioned isometries precisely 
bring us to these compactifications.
These, at low energies, reduce to compactifications of 
$10d$ supergravity on $S^2$ and $S^3$ times some torus, and therefore
connect Poincar\'{e} supergravities to anti-de Sitter ($adS$) supergravities. 
Compactifying eleven-dimensional supergravity on spheres instead of 
tori yields the latter. A famous example is the compactification of 
eleven-dimensional supergravity on $S^7$, which yields \cite{S7gsugra}
$N=8$ $adS_4$ gauged supergravity \cite{N8gaugesugra}. 
In other words, these transformations connect 
solutions of $10d$ (or of $11d$) supergravity 
that correspond to different compactifications. As such, they 
may connect solutions with different number of non-compact dimensions. From 
the point of view of M-theory, one may argue that all compactifications
of eleven-dimensional supergravity should be on an equal footing.
This suggests that the symmetry group of M-theory is actually larger 
than what is usually assumed. To obtain the full symmetry group 
one should also consider the various gauged supergravities.
The consistent picture that emerges from our discussion
of black-hole entropy strongly supports this point of view.
We shall, from now on in this article, use the term U-duality transformation
to denote the transformation that results from a combination 
of the usual ($R \leftrightarrow 1/R$) T-duality, of the S-duality of type-IIB
string theory and of the extra transformations that we mentioned above. 
We shall also freely uplift $10d$ results to eleven dimensions.

We shall argue that certain branes, and intersections thereof,
are U-dual to supersingleton representations of various anti-de Sitter 
groups. In this way we have a connection between our considerations 
and the usual D-brane picture.
In particular the branes $M2$, $M5$ and $D3$ are dual to the 
supersingleton representation of $adS_4$, $adS_7$, and $adS_5$, respectively. 
A complete list is given in Table 1 (see section 4). All the configurations
listed there (with the addition of a wave, in some cases) are 
dual to black holes in $4 \leq d \leq 9$. Essentially, the duality 
transformations map the black hole into the near-horizon geometry (with 
some global identifications). In the present context, however, this is 
{\it not} an approximation.

The picture emerging from our study is that the microscopic 
degrees of freedom reside in the intersection region of the various branes,
making up the black-hole configuration.\footnote{
We need not be 
in the weak string coupling limit for our considerations to be 
valid. In fact, we will always stay within the black-hole phase, 
where the string coupling is strong. Hence, by branes we mean the specific 
solutions of the low energy supergravity. In the case where the latter 
carry R--R charge, they are the long-distance description of D-branes.}
This picture is in harmony with results existing in the literature.
For $4d$ and $5d$ extremal black holes, 
described by a configuration of D-branes that has a one-dimensional 
intersection, the entropy can be obtained by treating the degrees of 
freedom as an ideal gas of bosons and fermions in a one-dimensional
compact space. Similar results (but with only qualitative agreement)
hold for near-extremal non-dilatonic black holes \cite{ts3}.
In that case as well, the microscopic description involves a $p$-dimensional 
theory, where $p$ is the spatial dimension of the intersection 
region. Notice, however, that the intersection region
is not a U-duality-invariant notion 
since the same black hole can result from different intersections.
For instance, the $5d$ black holes can be constructed 
by either the intersection of an M-theory membrane ($M2$) with 
an M-theory five-brane ($M5$) and a wave ($W$)
along the common direction, or from the intersection of three
membranes. In the former case the intersection is one-dimensional, 
i.e. over a string,  whereas in the latter it is zero-dimensional,
i.e. over a point. Let us emphasize that only U-duality-invariant quantities
of the original configuration may be studied in the U-dual 
formulation. The entropy of the black hole is such a quantity
and, therefore, can be computed in any dual configuration. 

This article is organized as follows. In sections 2 and 3 we  
concentrate on the $4d$ and $5d$ black holes. In particular,
in section 2 we show that $5d$ and
$4d$ non-extremal black holes are U-dual to configurations
that contain the BTZ black hole as the only non-compact part.
In section 3 we present our microscopic derivation of the 
Bekenstein--Hawking entropy formula. We follow Carlip's approach,
putting some emphasis on the unitarity issue of the underlying 
$SL(2, \IR)$ WZW model. In section 4 we discuss the duality 
between branes and supersingleton representations. In this way 
we provide a connection between our considerations and the D-brane picture. 
In section 5 we briefly discuss higher-dimensional black holes as well 
as intersections of branes (different from the ones discussed in section 2), 
which yield $4d$ and $5d$ black holes. We conclude in section 5.
Appendix A contains the eleven-dimensional supergravity
configurations that reduce, upon dimensional reduction along a 
compact direction, to the ten-dimensional solutions used in section 2.
Finally, in appendix B we show that higher than five-dimensional
black holes are not U-duals to configurations that contain the BTZ black hole.

\section{U-duality between non-extremal and \hfill\break 
         BTZ black holes}

We will show in this section that ten-dimensional configurations,
which upon dimensional reduction in an appropriate number of 
dimensions yield a $5d$ or a $4d$ black hole, can be mapped 
by a chain of dualities and a simple coordinate transformation 
into a configuration that has as the only non-compact part the 
BTZ black hole. In particular, the configuration that yields 
the $5d$ black hole will be mapped to the configuration
${\rm BTZ} \xx S^3 \times T^4$, and the one that yields the $4d$ black 
hole to ${\rm BTZ} \xx S^2 \xx T^5$. 
We will show that there is an exact 
CFT associated to each factor of the final configuration. 
For this to be true, it is crucial to carry along the gauge fields  
of the original configuration. After the dualities all the fields
acquire their canonical values so that each factor is independently
associated to a CFT. In this sense, our considerations also 
provide exact CFTs associated to $5d$ and $4d$ black holes.

The basic mechanism that allows one to map one black hole 
that is asymptotically flat into other that is asymptotically 
anti-de Sitter has been discussed in \cite{BPS}. 
Here we shall refine this discussion by showing that what was there called   
shift transformation, is actually a property of the 
plane-wave solution. Consider the following non-extremal 
plane-wave solution in $(D+1)$ dimensions 
\bea
ds^2 &=& - K^{-1}(r) f(r) dt^2 + K(r) \left(dx_1 + 
(K'{}^{-1}(r)-1+\tanh \a)dt \right)^2 \nonu
&&+ f^{-1}(r) dr^2 + r^2 d \O_{D-2}^2 ~ , \label{wave1}
\eea
where 
\bea 
&&K(r) = 1 + \frac{\m^{D-3} \sinh^2 \a}{r^{D-3}}~ , ~~~~ 
K'{}^{-1}(r) = 1 - \frac{\m^{D-3} \sinh \a \cosh \a}{r^{D-3}} K^{-1} ~ ,
\nonu
&&f(r)=1 - \frac{\m^{D-3}}{r^{D-3}}~ , ~~~~
r^2 = x_2^2 + \cdots + x_{D}^2 ~ .
\label{fdef}
\eea
The coordinate $x_1$ is assumed to be periodic,
with radius $R_1$, so that $(t, x_1)$ has the
topology of a cylinder.
The constant of the off-diagonal part is chosen such that this term 
vanishes at $r=\m$.  One may T-dualize in the $x_1$-direction
to obtain a solution that describes a non-extremal string. 
In this case, the off-diagonal part of the metric becomes 
the antisymmetric tensor of the new solution. Our choice of the 
constant in the off-diagonal part of (\ref{wave1}) ensures that
the latter is regular at the horizon \cite{hor1}.
We shall call the $r=\m$ surface horizon since, 
as we shall shortly see, the plane wave solution
when combined with certain other branes yields $5d$ and $4d$ black holes
solutions with an outer horizon at $r=\m$. 
The area of the latter for the solution (\ref{wave1}) is equal to 
\be \label{area}
A=2 \pi R_1 \cosh \a \m^{D-2} \O_{D-2}  ~ ,
\ee
where $\O_{D-2}$ denotes the volume of the unit $(D{-}2)${-}sphere.

Let us perform the following $SL(2, \IR)$ coordinate 
transformation that preserves the cylinder:
\be \label{transf}
\left(
\begin{array}{c}
t \\
x_1
\end{array}
\right) = 
\left(
\begin{array}{cc}
a & b  \\
0 & c
\end{array}
\right)
\left(
\begin{array}{c}
t' \\
x_1'
\end{array}
\right)~ .
\ee
Requiring that the transformed solution still be of the form (\ref{wave1})
and have vanishing off-diagonal part at $r=\m$ like (\ref{wave1}), 
and that the asymptotics be different, uniquely fixes $a, b, c$ to
\be
a = \cosh \a~ ,~~~~  b= - \exp (-\a)~ , ~~~~ c={1 \over \cosh \a} ~ .
\ee
One obtains\footnote{
In the extremal limit the transformation (\ref{transf}) and the
metric (\ref{wave2}) appear to be
singular. In this case, we have to rescale the coordinates $t'$ and $x_1'$ as
$t'\to t' \m^{D-3\over 2} $ and $x_1'\to x_1'/\m^{D-3\over 2}$. After taking 
the limit $\a\to \infty$ in such a way that the charge $Q=\m^{D-3} \sinh^2\a $
is kept fixed, we obtain a well-defined transformation (\ref{transf}) 
with $b=-1/(2a)$, $c=1/a$ and $a$ arbitrary. 
The metric (\ref{wave2}) has also a well-defined limit.}
(with the primes in $t'$ and $x_1'$ dropped)
\be
ds^2 = - \tilde{K}^{-1}(r) f(r) dt^2 + \tilde{K}(r) 
\left(dx_1 + (\tilde{K}^{-1}(r)-1)dt \right)^2 
+ f^{-1}(r) dr^2 + r^2 d \O_{D-2}^2 ~ ,
\label{wave2}
\ee
where now
\be \label{newha} 
\tilde{K}(r) = \frac{\m^{D-3}}{r^{D-3}} ~ .
\ee
Notice that the radius of $x_1$ is now equal to $R_1 \cosh \a$.  
We shall call the transformation (\ref{transf}) the shift 
transformation.\footnote{
The definition of the shift transformation is not the same 
as the one employed in \cite{BPS}. There the shift 
transformation acted on the fundamental string solution 
and it was a combination of the shift transformation as defined in 
this article and T-dualities.}
One easily checks that the area of the horizon (i.e. of the 
surface $r=\m$) 
is still equal to (\ref{area}).
We therefore conclude that the shift transformation does not change the
area of the horizon.

\subsection{$5d$ black holes}

Consider the solution of type-IIA supergravity that describes 
a non-extremal intersection\footnote{
All configurations studied in this article are built according to the 
rules of \cite{ts1}. In the extremal limit they 
are supersymmetric,
and they are constructed according to the intersection rules
based on the `no-force' condition \cite{inters}.}
of a solitonic five-brane ($NS5$) 
a fundamental string ($F1$) and wave ($W$) along one of the common 
directions. This configuration can be obtained from a solution of 
$11d$ supergravity as described in appendix A.
Let us wrap the $NS5$ in 
$(x_1, x_2, x_3, x_4, x_5)$, the  fundamental string ($F1$) in 
$x_1$ and put a  wave along $x_1$. 
The coordinates $x_i$, $i=1, \ldots, 5$, are assumed to be periodic, 
each with radius $R_i$. The metric, the dilaton 
and the antisymmetric tensor are given by
\bea \label{10d}
ds_{10}^2 &=&  
H_{f}^{-1} \left(-K^{-1} f dt^2 
+ K \left(d x_1 - (K'{}^{-1} -1)dt\right)^2\right) \nonu
&&+~ dx_2^2 + \cdots + dx_5^2 + H_{s5}(f^{-1} dr^2 + r^2 d\O_3^2) ~ ,
\eea 
and
\bea
&&e^{-2 \f} = H_{s5}^{-1} H_{f}~ ,~~~~ 
B_{01} = H_{f}'{}^{-1}-1 + \tanh \a_f~ , 
\nonu
&& H_{ijk} = {1 \over 2} \e_{ijkl} \pa_l H_{s5}'~ , ~~~~  
i,j,k,l=6, \ldots, 9 ~ ,
\label{100d}\\
&&  r^2 = x_6^2 + \cdots + x_9^2 ~ ,\nonumber
\eea
where the various harmonic function are given by 
(\ref{harm1}), with the identifications $H_T \to H_f$ and $H_F \to H_{s5}$.
One may also express the ``magnetic'' $NS5$ brane in terms 
of the dual ``electric'' field, 
\be  
B_{012345} = \coth \a_{s5} (H_{s5}^{-1}-1) + \tanh \a_{s5}~ .
\ee
The constant parts of the $B_{01}$ and $B_{012345}$ were chosen (using
a constant gauge transformation) 
such that the antisymmetric tensors are regular at the horizon.

Dimensionally reducing in $x_1, x_2, x_3, x_4, x_5$, one gets  
a $5d$ non-extremal black hole, whose metric in the Einstein frame is given by
\be
ds_{E, 5}^2 = - \l^{-2/3} f dt^2 + \l^{1/3} (f^{-1} dr^2 + r^2 d \O_3^2)~ ,
\label{BH5d}
\ee
where 
\be
\l = H_{s5} H_f K = \left(1 + \frac{\cq_{s5}}{r^2}\right)\left(1 
+ \frac{\cq_f}{r^2}\right)\left(1 + \frac{\cq_K}{r^2}\right)~ .
\ee
This black hole is charged with respect to the Kaluza-Klein gauge fields 
originating from the antisymmetric tensor fields and the metric. When all 
charges are set equal to zero one obtains the $5d$ Schwarzchild black hole.
The metric (\ref{BH5d}) 
has an outer horizon at $r=\m$ and an inner horizon at $r=0$. 
The Bekenstein--Hawking entropy may easily be calculated to be
\be \label{ent}
S={A_5 \over 4 G_N^{(5)}} = 
{1 \over 4} \frac{(2\pi)^5 R_1 R_2 R_3 R_4 R_5}{G_N^{(10)}} 
 \m^3 \O_3 \cosh \a_{s5} \cosh \a_f \cosh \a_K~ ,
\ee
where $\O_3$ is the volume of the unit 3-sphere and 
$G_N^{(5)}$ and $G_N^{(10)}$ are Newton's constant in five and ten dimensions, 
respectively.

We shall now show that this black hole is U-dual to the configuration
of the non-extremal BTZ black hole times a 3-sphere. 
This will be achieved by using the shift transformation and 
a series of dualities. Since neither dualities\footnote{
For T-dualities, this has been shown in \cite{hor1}. S-duality
leaves the Einstein metric invariant and, therefore, it does not change the 
area either.} nor the shift transformation change the area of the horizon, 
the Bekenstein--Hawking 
entropy of the resulting solution is the same as the one of the 
black hole we started from. The idea is to
dualize the fundamental string $F1$ and the $NS5$ into a wave, 
apply the shift transformation (\ref{transf}) and then 
return to the original configuration. One sequence of dualities that 
achieves that is, first, to perform $T_1S$ 
($T_i$ denotes T-duality along the $x_i$-direction,\footnote{
T-duality interchanges the type-IIA and type-IIB string theories. 
When restricted to the fields in the NS--NS sector, the T-duality
rules are the same as those of Buscher \cite{BUSCHER}. For the R--R background
fields the corresponding rules can be found in \cite{BHO}.}
and $S$ is the 
S-duality transformation of the type-IIB string theory). 
Then, the $NS5$ becomes a $D5$-brane, the wave a $D1$-brane and 
the fundamental string $F1$ a wave. So, we can use the shift transformation 
(\ref{transf}) in $(t, x_1)$ to change the harmonic function $H_f$, 
as in (\ref{newha}). In addition, the radius of $x_1$ is now equal 
to $R_1 \cosh \a_f$. Next, we perform $T_{1234}ST_1$. This 
yields a wave in $x_5$, a $D2$ in $(x_1, x_5)$ and a $D4$ in 
$(x_2, x_3, x_4, x_5)$. Now, we use the shift transformation
(\ref{transf}) in $(t, x_5)$ to change the harmonic function $H_{s5}$. 
The radius of $x_5$ also changes to $R_5 \cosh \a_{s5}$.
Finally, we return to the original configuration with the inverse dualities
(no shift transformations). The final result is given
by the metric in (\ref{10d}), but with\footnote{It 
is possible to obtain (\ref{hfhs5}) and (\ref{p01h}) below in a 
single step, by combining all preceding transformations into one 
element of the U-duality group. 
The same comment 
applies for the similar considerations in subsection 2.2.
Notice that the coordinate transformation (\ref{transf}) and the
subsequent $R \leftrightarrow 1/R$ duality, combine into a single
T-duality transformation along an isometry which is 
space-like everywhere, but at spatial infinity, where it becomes null.}
\be
H_{f} = \frac{\m^2}{r^2}~ ,~~~~~~~~ H_{s5} = \frac{\m^2}{r^2}~ ,
\label{hfhs5}
\ee    
and, in addition,
\bea
&& e^{-2 \f}=1~ ,~~~~~~ B_{01}=H_{f}^{-1}-1~ ,
\nonumber \\
&& H_{ijk}={1 \over 2}\e_{ijkl}\pa_l (H_{s5}{-}1)~ ~ ,~~~~~
i,j,k,l=6, \ldots, 9
\label{p01h} ~ .
\eea
Notice that the parameters $\a_f$ and $\a_{s5}$ associated to the charges
of the original fundamental string $F1$ and the solitonic five-brane $NS5$
appear only in the compactification radii of $x_1$ and $x_5$ respectively, 
and not on the background fields themselves.\footnote{In the extremal limit 
where $\mu\to 0$, $\a_K\to \infty$, $\a_{s5}\to \infty$ and 
$\a_f\to \infty$, with the corresponding charges kept fixed, we have to 
perform the contraction 
$t\to \m^2 t$, $x_1\to \m^2 x_1$, $x_i\to \m x_i$ ($i=2,3,4,5$) and 
$\a' \to \m^2 \a'$. Then (\ref{10d}) and (\ref{100d})
have well-defined limits, and similarly for (\ref{6sol}) and (\ref{pbtp})
below.}

Dimensionally reducing along $x_2, x_3, x_4, x_5$ we find
\be
ds^2_{E,6}=ds_{BTZ}^2 + l^2 d \O_3^2 ~ ,
\label{6sol}
\ee
where 
\be
ds_{BTZ}^2= -\frac{(\r^2 - \r_+^2)(\r^2 - \r_-^2)}{l^2 \r^2} dt^2
+ \r^2 (d \varphi - \frac{\r_+ \r_-}{l \r^2} dt)^2 + 
\frac{l^2 \r^2}{(\r^2 - \r_+^2)(\r^2 - \r_-^2)} d \r^2 ~ 
\label{dsbtz}
\ee
is the metric of the non-extremal BTZ black hole 
in a space with cosmological constant 
$\L=-1/l^2$, with inner horizon at $\r=\r_-$ and  outer horizon at $\r=\r_+$.  
The mass and the angular momentum the BTZ black hole are equal to 
$M=(\r_+^2 + \r_-^2)/l^2$ and $J=2 \r_+ \r_-/l$. 
In terms of the original variables:
\bea
&& 
l =\m~ ,~~~~~ \varphi={x_1 \over l}~ , ~~~~~ \r^2 = r^2 + l^2 \sinh^2 \a_K~ ,
\nonumber\\
&& \r_+^2 = l^2 \cosh^2 \a_K~ ,~~~~~~ \r_-^2 = l^2 \sinh^2 \a_K ~ .
\eea  
In addition,
\be
\f =0~ ,~~~~~  B_{t \varphi}=(\r^2 - \r_+^2)/l~ ,~~~~~ H =l^2 \e_3~ ,
\label{pbtp}
\ee
where $\e_3$ is the volume form element of the unit 3-sphere.
Therefore, the metric (\ref{6sol}) 
describes a space that is a product of a 3-sphere of radius $l$ and of 
a non-extremal BTZ black hole. 
Notice that the BTZ and the sphere part are completely decoupled. 
Also all fields have their canonical value, so that both  
are separately exact classical solutions of string theory, i.e. 
there is an exact CFT associated to each of them. 
For the BTZ black hole the CFT corresponds to an orbifold of the WZW model
based on $SL(2,\IR)$ \cite{hor2,kaloper}, whereas for $S^3$ and the associated
antisymmetric tensor with field strength $H$, given in (\ref{pbtp}), the 
appropriate CFT description is in terms of the $SO(3)$ WZW model.

We can now calculate the entropy of the resulting black hole.
The area of the horizon is equal to 
\be
A_3= 2 \pi {R_1 \cosh \a_f \over \m} \m \cosh \a_K~ ,
\ee
whereas Newton's constant is given by 
\be
G_N^{(3)} = { G^{(10)}_N \over ((2 \p)^4 R_2 R_3 R_4 R_5 \cosh \a_{s5}) 
(\m^3 \O_3)} ~ .
\label{new3}
\ee
It follows that $S=A_3/(4 G_N^{(3)})$ equals (\ref{ent}), i.e. 
the Bekenstein--Hawking entropy of the final configuration is equal to the 
one of the original $5d$ black hole. Notice that the Newton constant in 
(\ref{new3}) contains the parameter $\a_{s5}$, i.e. carries information on
the charge of the original $NS5$ five-brane.

\subsection{$4d$ black holes}

Consider the solution of type-IIA supergravity that describes a
non-extremal intersection of a $D2$ brane in $(x_1, x_2)$, 
a $D6$ brane in $(x_1, x_2, x_3, x_4, x_5, x_6)$,
a solitonic five-brane $NS5$ in $(x_1, x_3, x_4, x_5, x_6)$ with a 
wave along $x_1$. The eleven-dimensional origin of this solution
is described in appendix A. The coordinates $x_i, i=1, \dots, 6$, 
are assumed to be periodic, each with radius $R_i$.
The metric, the dilaton and the antisymmetric tensors are given by
\bea \label{4d}
d s_{10}^2 &=& (H_6 H_2)^{-1/2} 
\left(- K^{-1} f d t^2 + K (d x_1 + (K'{}^{-1} -1)dt)^2\right) 
\nonu
&&+H_{s5}(H_6 H_2)^{-1/2} d x_2^2
+H_6^{-1/2}H_2^{1/2}(dx_3^2+dx_4^2+dx_5^2+d x_6^2) \\
&&+H_{s5}(H_6 H_2)^{1/2}(f^{-1}dr^2+r^2d\O_2^2) ~ ,
\nonumber 
\eea
and 
\bea 
&&e^{- 2 \f}=H_{s5}^{-1} H_6^{3/2} H_2^{-1/2}~ ,~~~~
H_{2ij}={1 \over 2} \e_{ijk} \pa_k H'_{s5}~, ~~~ i,j,k=7,8,9 ~ ,
\nonu
&&(dA)_{ij}={1 \over 2} \e_{ijk} \pa_k H_6'~ ,~~~~
C_{012}=\coth\a_2 (H_2^{-1}{-}1){+}\tanh\a_2 ~ ,
\eea
where the various harmonic functions are given in (\ref{harm2}) 
of appendix A (but renamed as $H_{F1} \to H_{s5}$,
$H_{F2} \to H_{2}$ and $H_{F3} \to H_{6}$).

Upon dimensional reduction in $x_1, x_2, x_3, x_4, x_5, x_6$, one obtains 
a charged $4d$ non-extremal black hole\footnote{Extremal $4d$ black hole 
solutions embedded in  eleven-dimensional supergravity where constructed 
in \cite{kletse4d}.
In particular, these authors showed that a configuration of three intersecting 
five-branes with a wave along a common string and another
one of two membranes and two five-branes reduce, upon compactification to four
dimensions, to the extremal limit of (\ref{BH4d}).} 
with metric in the Einstein frame 
given by 
\be
ds^2_{E,4}= - \l^{-1/2} f dt^2 + \l^{1/2} (f^{-1} dr^2 + r^2 d \O_2^2)~ ,
\label{BH4d}
\ee
where
\be
\l= H_{s5} H_6 H_2 K  = 
\left(1+ { \cq_{s5}\over r}\right)\left(1+ {\cq_6 \over r}\right)
\left(1+ {\cq_2 \over r}\right)\left(1+ {\cq_K \over r}\right)~ .
\ee
The antisymmetric tensor fields and the off-diagonal part of the
metric give rise to gauge fields under which this solution is charged.
The usual $4d$ Schwarzchild black hole is obtained by setting all 
charges equal to zero.
The metric (\ref{BH4d})
has an outer horizon at $r=\m$ and an inner horizon at $r=0$. 
The Bekenstein--Hawking entropy may easily be calculated to be 
\be \label{ent1}
S={A_4 \over 4 G_N^{(4)}} = 
{1 \over 4} \frac{(2\pi)^6 R_1 R_2 R_3 R_4 R_5 R_6}{G_N^{(10)}} 
 \m^2 \O_2 \cosh \a_{s5} \cosh \a_6 \cosh \a_2 \cosh \a_K ~ ,
\ee
where $\O_2$ is the volume of the unit 2-sphere and 
$G_N^{(4)}$ is Newton's constant in four dimensions. 

In order to show that this black hole is dual to the BTZ one, we use the 
same strategy as before. We dualize the solution in such a way 
that each brane becomes
a wave, then we apply the shift transformation, and we finally return
to the original configuration with the inverse dualities. For instance, the 
chain of dualities $T_1ST_{3456}ST_1$ converts the $NS5$ into a wave.
In order to convert $D2$ into a wave one may use the dualities 
(starting from the original configuration) $T_2ST_1$. 
Finally, the $D6$ may be converted to $D2$ by $T_{3456}$. Then, 
one may use the same dualities as in the previous case. The combined 
effect of these dualities is to change the radius of $x_1$ to 
$R_1 \cosh\a_{s5}$, the radius of $x_2$ to $R_2 \cosh\a_6 \cosh\a_2$, 
the harmonic functions to
\be 
H_{s5}=H_6=H_2={\m \over r} ~ ,
\ee
and the fields to
\bea
&&e^{-2\f}=1~ , ~~~~~ C_{012}=H_2^{-1}-1~ ,
\nonu
&&H_{2ij}={1 \over 2} \e_{ijk} \pa_k H_{s5}~ ,~~~ 
(dA)_{ij}={1 \over 2} \e_{ijk} \pa_k H_6~ ,~~ i,j,k=7,8,9~ .
\eea
Similarly to the
case of subsection 2.1, the parameters $\a_2$, $\a_6$ and $\a_{s5}$
associated with the charges of the original $D2$, $D6$ and $NS5$ respectively,
appear only in the compactification radii of $x_1$ and $x_2$, but not in
the background fields themselves.\footnote{
In the extremal limit we have to perform a contraction similar to the one
described in footnote 9.}

After dimensional reduction in $x_2, x_3, x_4, x_5, x_6$, one gets
\bea 
ds^2_{E,5}=ds_{BTZ}^2 + \m^2 d \O_2^2 ~ ,
\label{5sol}
\eea
where 
\bea
&& l = 2 \m~ ,~~~~ \varphi={x_1 \over l}~ , ~~~~
\r^2 = 2 l r + l^2 \sinh^2 \a_K ~ ,  \nonumber \\
&& \r_+^2 = l^2 \cosh^2 \a_K~ ,~~~~~ \r_-^2 = l^2 \sinh^2 \a_K ~ .
\eea  
In addition,
\be
\f =0~ ,~~~~  B_{t \varphi}=(\r^2 - \r_+^2)/l~ ,~~~~ 
F = \m \e_2 ~ , 
\ee
where $F$ represents the $U(1)$ field strengths and 
$\e_2$ is the volume form element of the unit 2-sphere.
As in 
the case of the five-dimensional black hole we also see that the BTZ black 
hole and the 2-sphere decouple completely. 
We also note that the second term in (\ref{5sol}), representing the 2-sphere
with the associated gauge field we have mentioned, corresponds to 
the monopole CFT of \cite{GPSmonopole}. Equivalently, it can also be
viewed as a dimensionally reduced  $SO(3)$ WZW model along one of the 
Euler angles parametrizing the $SO(3)$ group element.

One may calculate the entropy of the final configuration. The result is 
in agreement with (\ref{ent1}). As in the five-dimensional case,
the volume of the sphere as well as certain parameters associated
with charges of the original configuration (\ref{4d}), enter via
the three-dimensional Newton constant.   

\section{Counting microscopic states}

In this section we briefly review Carlip's derivation of the 
Bekenstein--Hawking entropy formula for the BTZ black hole. 
The basic idea is that
only quantum states leaving on the horizon of the black hole are relevant 
in such computation, whereas those in the bulk are irrelevant.
Since the horizon represents the end of the world for an outside 
observer, it is treated as a surface boundary.
This is in principle applicable in any number of dimensions, and
the problem is to be able to separate the boundary from the bulk degrees
of freedom and subsequently to quantize them. 
This is a formidable task by itself
in more than three space-time dimensions, and we know of no solution to date. 
However, in $(2+1)$ dimensions the
problem is trivially solved since there are no
bulk degrees of freedom at all. Moreover, as we have seen in section 2
this is enough
for our purposes, since we have mapped the problem of counting microscopic 
states for the $4d$ and $5d$ black holes into the corresponding problem for the
$3d$ BTZ black hole. The topological character of $(2+1)$-dimensional
gravity is manifest in its Chern--Simons formulation \cite{town1, wit}.
In the presence of a non-vanishing cosmological constant the action can be 
written as\footnote{
We only give the bosonic part. The full supersymmetric version
has also a Chern--Simons form, but in superspace \cite{town1}. 
In principle, one should also keep the fermions in the derivation
of the boundary action. The latter, however, 
at least in the limit of small cosmological constant,
have subleading contribution to the entropy. Nevertheless, it will be 
useful to repeat the computation by including the fermions as well.}
\be
S = S_{CS}(A) - S_{CS}(\tilde A) ~ ,
\label{2p1gr}
\ee 
where\footnote{Our normalizations are compatible with the representation
$T_0 = i\s_3/2$, $T_1= \s_1/2$, $T_2= \s_2/2$ 
for the $SL(2,\IR)$ generators, and $\Tr$ is the matrix trace.}
\be
S_{CS}(A) = {k\over 8\pi} \int_M \Tr\left(A \wedge d A 
+  {2\over 3} A\wedge A \wedge
A \right)      ~ 
\label{scs}
\ee
represents the Chern--Simons actions for some manifold $M$ and similarly for 
$S_{CS}(\tilde A)$. The gauge connections are in the Lie algebra of $SO(1,2)$
and are given in terms of the spin connection $\o^a$ and triad $e^a$
1-forms as 
\be
A^a = \o^a - {e^a\over l} ~ ,~~~~~~~ \tilde A^a = \o^a + {e^a\over l} ~ ,
\label{atila}
\ee
where $a=0,1,2$.
We will denote
$A^\pm_\mu = A^1_\mu \pm A^0_\mu$ and similarly for
$\tilde A^\pm_\mu$. 
The constants $k$ and $l$ are related by $k={l\over 4 G^{(3)}_N }$.\footnote{ 
Notice that, since the Newton constant depends on various charges 
(as follows from our discussion in the previous section), so 
does $k$. This resonates with the idea of the string-tension renormalization
employed in \cite{finn}.}
It is well known that,
if the manifold $M$ has no boundary, a Chern--Simons theory in 
$(2+1)$ dimensions is a topological field theory.
However, if there is a non-trivial boundary $\partial M$,
then the variational problem of pure Chern--Simons is not well defined unless 
we specify the  boundary conditions and add a boundary-action term $S_B$. 
This is the case of interest to us, where the non-trivial 
boundary will be identified with the (apparent) horizon of the 
$(2+1)$-dimensional BTZ black hole. This, in turn, is a guideline for fixing 
the appropriate boundary conditions. We will briefly repeat the arguments of
\cite{carlip} (see also \cite{carlipother}
for a systematic general discussion of boundary 
conditions and edge states in gravity).
We change coordinates from $(t,\r, \phi)\to (u,v,\phi)$, where $u$ and $v$
are light-cone coordinates (the precise relation can be read off by 
comparing (\ref{dsbtz}) and eq. (3.1) of the first article in 
\cite{carlip}). 
Consider the boundary, with the topology of a cylinder, parametrized 
by the angular variable $\phi$ and the non-compact variable $v$.
Keeping $A^2_\phi$, $A^+_\phi$, 
$A^+_v$, as well as their tilded counterparts,
fixed in the boundary, requires that the
action $S_B$ be given by 
${-k\over 8\pi} \int_{\partial M} dt d\phi \left(A_\phi^2 A_v^2 + \half
(A^+_\phi A^-_v - A^-_\phi A^+_v) \right)$
minus a similar term with
$A$'s replaced by $\tilde A$'s. The total action is given by the sum of
(\ref{2p1gr}) and $S_B$ and, as a result, the variational problem is
now well defined.
The relevant degrees of freedom in
the boundary are isolated by parametrizing $A= g\inv A_f g + g\inv d g$,
where $A_f$ is a fixed gauge connection in the boundary, and similarly 
for $\tilde A$. Then, 
quite generally, it can be shown that the relevant induced action in the 
boundary is the sum of two WZW actions for $SL(2,\IR)$ with opposite levels:
\be
S_B = kI_0(g) - k I_0(\tilde g)  ~ ,  
\label{sbou}
\ee
As we have already mentioned, since the horizon of the BTZ black hole 
at $\rho = \rho_+$ 
(which in the new coordinates is located at $u=0$)
is a null surface, we should demand that
the boundary $\partial M$ be a null surface as well.
It was shown in \cite{carlip} that the appropriate
boundary conditions that achieve this are
$A^+_\phi = A^+_v = \tilde A^+_\phi = \tilde A^+_v =0$.
Moreover, we should demand that the circumference of the boundary
be the same as that of the BTZ black hole,
namely $2\pi \rho_+$.\footnote{The actual circumference 
is $2 \pi \d \r_+$, where $\d= R_1 \cosh \a_f/\m$ or
$\d=R_1 \cosh \a_{s5}/(2\m)$ depending on whether the BTZ black hole 
corresponds to the $5d$ or to the $4d$ black hole. Rescaling 
$t\to t \d^2$, $l \to l \d$, $\a' \to \a'\d^2$ and $\phi \to \phi \d$ 
effectively sets $\d=1$. Since, in three dimensions,
$G^{(3)}_N \sim \a'^{1/2}$ the Newton constant is accordingly rescaled. 
It is possible to perform the counting of states using the original 
parameters, but we prefer the rescaled ones, 
thus keeping contact with the original computation by Carlip.}
Then, a natural boundary
condition, which is also in agreement with the metric (\ref{dsbtz}),
is $e^2_\phi = \rho_+$.
What remains is to choose a 
boundary condition for $\o^2_\phi$.
As there is no physical principle that has not been met 
at this point, we leave its boundary value undetermined for the moment.
The aforementioned 
boundary conditions are not invariant under the full two-dimensional 
group of diffeomorphisms but only under rigid translations of the 
angular variable $\phi$. Hence, we must impose on the Hilbert space of 
(\ref{sbou}) the constraint
\be
L^{total}_0 = L_0 + \tilde L_0 = 0 ~ ,
\label{lolo}
\ee
where $L_0$ and $\tilde L_0$ are the zero modes of the Virasoro generators
corresponding to the affine algebras for $A^a_\phi$ and $\tilde A^a_\phi$
in (\ref{sbou}).
The expectation value of $L^{total}_0$, in a Hilbert space state of total 
level $N$, assumes the form
\be
L^{total}_0 = N +  {C_{sl(2,\IR)}\over k-2}  -
{\tilde C_{sl(2,\IR)}\over k+2} = 0 ~ ,
\label{ltot}
\ee
with the Casimir operators given by
\bea
C_{sl(2,\IR)} & = &  (A^2_0)^2 + \half (A^+_0 A^-_0 + A^-_0 A^+_0 ) ~ = ~ 
- j (j+1) ~ ,
\nonumber \\
\tilde C_{sl(2,\IR)} & = & (\tilde  A^2_0)^2 
+ \half (\tilde A^+_0 \tilde A^-_0 
+ \tilde A^-_0 \tilde A^+_0 ) ~ = ~ - \tilde j (\tilde j+1) ~ ,
\label{casii}
\eea
where $A^a_0$, $a=2,+,-$, are the zero modes in a Fourier series expansion
of the gauge connection $A^a_\phi$,
i.e. $A^a_\phi= {1\over k} \sum_{n=0}^\infty A^a_n e^{i n \phi}$ and obey
the Lie algebra $sl(2,\IR)$. A similar expression holds 
for $\tilde A^a_\phi$ as well. 
Also $j$ and $\tilde j$ label the representation of 
$sl(2,\IR) \otimes sl(2,\IR)$. Recall that, we have imposed on the boundary 
that $A^+_\phi = \tilde A^+_\phi =0$. 
Hence, the Casimir operators in (\ref{casii})
are positive-definite. 
As we shall see, this fact and simple thermodynamical
considerations, allow only for the principal series representation.
Using the boundary 
condition $e^2_\phi = \rho_+$ and the definition (\ref{atila}) we may express
the zero modes as 
\be
A_0^2 = k \left(\o -{\rho_+ \over l}\right) ~ ,~~~~~ 
\tilde A_0^2 = k \left(\o + {\rho_+ \over l}\right) ~ ,
\label{zerer}
\ee
where $\o$ denotes the zero mode of $\o^2_\phi$
and encodes the remaining freedom in choosing boundary conditions.
Then, using (\ref{ltot}), we find that 
\be
N =  {k^2\over k+2} \left(\o + {\rho_+\over l}\right)^2 
-  {k^2\over k - 2} \left(\o - {\rho_+\over l}\right)^2 ~ .
\label{ntot}
\ee
In the thermodynamic limit the configurations with maximum number of states 
dominate. Hence, we should maximize $N$
with respect to $\o$. It can be easily shown that the maximum value of $N$ is
reached for $\o= \o_m \equiv { k \rho_+ \over 2 l}$ and that it is given by
\be
N_m =  { k^2 \rho_+^2 \over  l^2} ~ .
\label{ntomax}
\ee
Finally, the entropy is computed by using the fact that for a 
CFT with central charge $c$ the number of states behaves asymptotically 
at large levels $N$ as \cite{cardy}
\be
n(N) \approx \exp\left ( 2 \pi \sqrt {{N\over 6} c} \right) ~ , ~~~~~~ 
N\gg 1 ~ .
\label{assymp}
\ee
Using the leading order in $k$ value for the central charge,
i.e. $c\approx 6$, one computes the entropy to be \cite{carlip}
\be
S = \ln n(N_m) \approx {2 \pi \rho_+ \over 4 G^{(3)}_N} ~ .
\label{entt}
\ee
This is precisely the Bekenstein--Hawking entropy formula for the BTZ black 
hole.

We next prove that, due to boundary conditions, only principal series
representations of $sl(2,\IR)$ are allowed in 
the thermodynamic limit, in which $N_m\gg 1$. 
This limit is what 
one intuitively 
expects from a physical point of view, but it can also be established by 
requiring that for the statistical description to be valid the 
condition $\big | {\partial T \over \partial M}\big|_J \ll 1$ should be 
fulfilled \cite{valista}, where $T=(\rho_+^2- \rho_-^2)/(2\pi \rho_+ l^2)$ 
is the temperature of the black hole. 
In our case we have explicitly
\be
\bigg | {\partial T \over \partial M}\bigg |_J ={1\over \pi}
 {N_m^2 + 24 J^2 k^2 \over N_m^2 - 8 J^2 k^2} 
{1\over \sqrt{N_m}} \ll 1 ~ ,
\label{sfsg}
\ee
which implies that $N_m \gg J k $.
Due to boundary conditions, (\ref{casii}) reduces to $j(j+1) + (A_0^2)^2=0$,
where $A_0^2$ is given by (\ref{zerer}) (computed for $\o=\o_m$).
This algebraic equation is solved for $j$ to give 
\be
j = -\half \pm \sqrt{{1\over 4} - N_m (k-2)^2}~ ,
\label{jsol}
\ee
where we have used (\ref{ntomax}). It is clear that 
the discrete series representations for which $j+1 >0$ and $j\in Z+\half$ (or
$j\in \IR$ if we consider the universal cover of $sl(2,\IR)$) and 
the supplementary series for which $j=-1/2 + s $, $0 < |s| < \half $,
are not allowed if $N_m \gg 1$, since then $j$ becomes complex.
However, this is precisely what is needed for 
the continuous series representation to be allowed, since in this case
$j=-1/2 + i \s $, $\s \in \IR$. Identifying the latter expression with the
one in (\ref{jsol}), after it is rewritten so that it is valid 
for large $N_m$, we obtain 
\be
\s^2 = N_m (k-2)^2 -{1\over 4}~ .
\label{s2p}
\ee
For $\tilde j$ the corresponding $\tilde \s$ is given by an expression 
similar to (\ref{s2p}), but with $k$ replaced by $-k$.
Clearly, the right-hand side of (\ref{s2p}) is positive 
for a sufficiently large number of states $N_m$, i.e. the principal series 
representation is allowed.

Our final comment concerns the issue of unitarity in WZW models 
based on non-compact groups. In general, this is still an
unsolved problem (for earlier work on the subject, see
\cite{noncomp,char,grifher}), 
but in the case of the $SL(2,\IR)$ WZW model  
it has been argued that a consistent, unitary theory, can be obtained 
by restricting to highest-weight states belonging to the principal series
representation \cite{bars}. 
In this case the current algebra character formula
is the same as that of a theory of three free bosons \cite{char}.
However, the construction of modular invariants is subtle,
essentially because states in the Verma module corresponding to 
the principal series representation do not form a closed set under the fusion
rules \cite{grifher}. 
In addition, if we try to construct modular invariants by using only
principal series representations, we would need 
to obtain the appropriate measure of integration 
over all $j=-\half + i \s$.
Notice, however, that since the boundary conditions break the
two-dimensional diffeomorphisms it may not be necessary to have a modular 
invariant formulation; only the norm of the microstates is required 
to be positive-definite. 
It would be important to reexamine these and related issues in view of 
the great relevance of the $SL(2,\IR)$ WZW model            
in black-hole physics we have uncovered.


\section{Connection with D-branes}

Since we want to compare our counting of microscopic black-hole states 
with the counting using D-branes, 
let us consider the extremal case where the latter is valid.
In the D-brane picture, one constructs a configuration of D-branes
that carries the same quantum numbers as the corresponding 
black hole. Counting the degeneracy of this configuration 
yields the number of microstates. When we uplift it
to M-theory it becomes an intersection  
of membranes $M2$, five-branes $M5$ and plane-wave $W$ solutions.

The effect of the shift transformation on the M-branes and 
on intersections of them has been studied in \cite{BPS}.
The result is that certain branes and intersections thereof
are mapped into spaces that are locally isometric to 
spaces of the form $adS_k \xx E^l \xx S^m$, where $adS_k$ denotes 
the $k$-dimensional anti-de Sitter space, $E^l$ denotes 
the $l$-dimensional Euclidean space and $S^m$ is the $m$-dimensional
sphere. We tabulate these results below.
We also give the result for the $D3$-brane.
Similar results hold for the rest of the branes, but only when 
they are expressed in the ``dual $Dp$-frame'', i.e. the metric in which the
curvature and the $(8{-}p)$-form field strength appear in the action
with the same power of the dilaton \cite{BPS2}.
In all cases, in order to arrive at the dual configuration 
one needs a number of compact isometries. This yields the space 
indicated in the second column of the table with some global 
identifications. For instance, the $adS_3$ appearing below is 
more properly viewed as an extremal BTZ black hole (with $J \neq 0$
only if a plane wave is added to the corresponding 
configuration in the left column).
\newpage

\begin{center}
\vspace{.2cm}
\centerline{\bf Table 1}
\vspace{.2cm}
\begin{tabular}{|c|c|}
\hline
$M5$ & $adS_7 \xx S^4$\\ \hline
$M2$ & $adS_4 \xx S^7$\\ \hline
$D3$ & $adS_5 \xx S^5$\\ \hline
$M2\perp M5$ & $adS_3\times E^5\times S^3$\\ \hline
$M5\perp M5\perp M5$ & $adS_3\times E^6\times S^2$\\ \hline
$M2\perp M2\perp M2$ & $adS_2\times E^6\times S^3$\\ \hline
$M2\perp M2\perp M5\perp M5$ & $adS_2\times E^7 \times S^2$\\
\hline
\end{tabular}
\end{center}
\vspace{.2cm}

It is rather remarkable that these considerations distinguish 
branes and intersections that we already know to
play a distinguished r\^{o}le for other reasons. 
For instance, from these configurations
(with the addition of a wave in some cases) one can obtain 
black-hole solutions in $4 \leq d \leq 9$ upon dimensional 
reduction. 

Since after the duality the asymptotic geometry has changed,
the degrees of freedom should organize themselves into 
representations of the appropriate anti-de Sitter group. 
The latter has some representations, the so-called singleton 
representations, that  have no Poincar\'{e} analogue.\footnote{These 
representations, for the case of $adS_4$, were
discovered by Dirac \cite{Dirac} and named singletons by
C. Fronsdal \cite{Fronsdal}.}
They have appeared in studies of spontaneous 
compactifications of eleven-dimensional supergravity 
on spheres. In particular, the fields of the 
supersingleton representation appear as coefficients in the 
harmonic expansion of the eleven-dimensional fields 
on the corresponding sphere. A crucial property is that the singleton
multiplets can be gauged away everywhere, 
except in the boundary of the anti-de Sitter space \cite{gauge}.   
In particular, it has been argued in the past that  
the singleton representations of $adS_4$, $adS_7$, $adS_5$ and $adS_3$
correspond to membranes \cite{m2}, five-branes \cite{NST,m5}, self-dual 
threebranes \cite{NST,m5} and strings \cite{f1}, respectively.
It has actually been shown that, in all cases, the world-volume
fields of the corresponding $p$-brane form a supersingleton multiplet.  
We, therefore, conclude that the membrane 
$M2$, the five-brane $M5$, the self-dual 
threebrane $D3$, as well as strings,
are U-dual to supersingletons.
Looking back to Table 1, we see that the anti-de Sitter 
spaces appearing there, are precisely the ones we just discussed, 
with one exception, the $adS_2$ space. The boundary of $adS_2$ is simply 
a point. Thus, one deals with quantum mechanics instead of quantum 
field theory. It is very tantalizing to identify the theory on the 
boundary with $D0$ branes. This might yield a connection with 
M(atrix) theory. However, the $D0$ solution factorizes
as $adS_2 \xx S^8$ only in the ``dual-8 frame''. So, it is not 
clear whether or not such an identification is correct.   

What is important is that, 
precisely as in our discussion of the counting of states in 
section 3, would-be gauge degrees of freedom become dynamical at the boundary.
Let us consider, for concreteness, the case of the extremal $5d$ black hole.
The M-theory configuration is the intersection of an 
$M5$ wrapped in $(x_1, x_2, x_3, x_4, x_5)$, an $M2$ 
wrapped in $(x_1, x_{10})$ with a wave along $x_1$. The $5d$ black hole 
arises after a dimensional reduction along $x_1, x_2, x_3, x_4, x_5, x_{10}$.
Let us first consider the effect of the shift transformation to 
each brane separately (i.e. consider a configuration with only that 
brane). The $M5$ becomes the singleton representation of $adS_7$.
The anti-de Sitter space has coordinates $t, x_1, \ldots, x_5, r$. 
The coordinate $r$ used to be 
the radius of the transverse space. The five-brane is represented by gauge
degrees of freedom everywhere except at the boundary.
Studying the five-brane dynamics is equivalent to studying 
the supersingleton dynamics of $adS_7$. In a similar fashion,
the membrane becomes the singleton representation
of $adS_4$ (with coordinates $t, x_1, x_{10}, r$).
After superposition the effects of the two branes cancel each other 
in the relative transverse directions. We end up 
with $adS_3 \xx E^5 \xx S^3$, where the $adS_3$ part is along 
the common world-volume directions. 
It follows that the latter contains
gauge degrees of freedom that become dynamical at the boundary.
These correspond to the singleton representation of $adS_3$,
which can be interpreted as a string \cite{f1}. Thus, we find
a string living on the world-volume of the five-brane \cite{verl}. 
Notice that the anti-de Sitter group $SO(d-1,2)$ coincides 
with the conformal group in one dimension lower. 
Therefore, one ends up with a conformal field theory 
on the boundary. Since we are considering extremal black holes,
the theory at the boundary is also supersymmetric. 
After the addition of the wave along $x_1$, the $adS_3$ becomes a 
massive extremal BTZ black hole. 
These are precisely the degrees of freedom we have counted
in section 3. A similar interpretation holds also 
for the $4d$ black hole.   

Notice that the non-extremal black holes
result from the non-extremal intersection of extremal branes 
and not from the intersection of non-extremal branes. 
In other words, they can be viewed as 
non-extremal ``bound-state'' configurations \cite{ts1}. 
This means that one still has the interpretation 
of each brane as a singleton representation of the corresponding
anti-de Sitter group. Therefore, the above discussion still applies.

\section{Higher-dimensional black holes and \hfill\break
         further comments}

Let us briefly discuss higher-dimensional 
($6 \leq d \leq 9$) black holes.
These cases are more complicated, since they are not connected 
to three-dimensional black holes. A direct proof that the BTZ black 
hole cannot appear in U-dual configurations of these black holes
is given in appendix B. Already from the discussion of the 
previous section, however, it follows that the higher than five-dimensional
black holes are associated with higher than three-dimensional theories.
The $9d$ black holes can be obtained from the non-extremal intersection
of $M2$ with a wave, $7d$ black holes from the intersection
of $D3$ with a wave, and $6d$ black holes from the intersection
of $M5$ with a wave \cite{ts1}. Hence, these black holes\footnote{
The $8d$ black hole does not seem to be on an equal footing with 
the rest. One may obtain $8d$ black holes from 
a configuration of an $M2$ brane with a wave that has
an extra isometry along which one may dimensionally reduce.
This implies, however, that the corresponding sphere does not decouple.} 
are associated  with the first three entries of Table 1. 
It follows that in order to understand them one would need 
to understand the boundary field theories of $adS_4$, $adS_5$ and $adS_7$, 
respectively. Our considerations also imply that the metrics (supplied
with the appropriate antisymmetric tensor fields),
after we remove the part corresponding to the sphere,
describe solutions of gauged
supergravities in four, five and seven dimensions.
Presumably, they are black-holes solutions,
but this question deserves further study.

The fourth and fifth entries of Table 1,
when supplemented with waves, correspond 
to the $5d$ and $4d$  black holes we discussed in section 2. Closely 
related are the configurations of the last two entries of Table 1: 
they also correspond to $5d$ and $4d$ black holes. In these cases the 
non-compact part is a two-dimensional configuration, instead of the 
three-dimensional BTZ black hole. This can be thought of as the dimensionally 
reduced BTZ black hole along a compact direction. 
For both cases, there is an associated exact CFT. 
In particular, the last entry of Table 1,
after dimensional reduction along the directions of $E^7$,
corresponds to the $4d$ configuration $adS_2 \times S^2$. 
This is the Bertotti--Robertson metric, which (with appropriate gauge fields)
corresponds to an exact classical solution of string theory \cite{bertotti}.
Notice that
the last three entries of Table 1 can be obtained from ${\rm BTZ} \xx S^3$ 
after we dimensionally reduce
along appropriate Euler angles parametrizing the corresponding group elements.
Reducing the BTZ part one obtains the $adS_2$ black hole, whereas
reducing the $S^3$ part one obtains $S^2$. In all cases the CFT description is
in terms of the original one for ${\rm BTZ} \xx S^3$
(the various gauge fields are important for this). 

One may also consider the first five entries of Table 1 without the 
addition of a wave. These are non-dilatonic black branes whose thermodynamic
properties were studied in \cite{ts3}.
All of them have zero entropy in the extremal limit.\footnote{This can 
easily be seen from (\ref{ent}) and (\ref{ent1}) by first setting $\a_K=0$,
i.e. no wave, and then going to the extremal limit $\mu\to 0$ with
the charges 
($\cq_{s5}$, $\cq_f$) and ($\cq_{s5}$, $\cq_2$, $\cq_6$), respectively,
kept fixed.}
Near-extremality, however, their entropy has the same form as  
the entropy of an ideal gas of massless particles. For $M5$, $M2$ and $D3$
the entropy behaves, as a function of the temperature $T$,
as $S_p \sim T^p$, where $p=5, 2$ and $3$, respectively.
This is the scaling behaviour of the entropy of an ideal
gas of massless particles 
in $p$ spatial dimensions. For the fourth and fifth entries one gets
$S \sim T^1$, i.e. a string-like form. 
It is now easy to 
understand these results. From our previous discussion we 
know that the degrees of freedom that account for the Bekenstein--Hawking
entropy live on the boundary of the 
corresponding anti-de Sitter space. The latter has precisely
the right dimension in each case. Near extremality,
the degrees of freedom interact only weakly, and therefore one may 
associate to them a gas of free particles. Away from extremality, when
the various interactions are turned on, 
full knowledge of the boundary dynamics is required.
  

\section{Conclusions}

We have presented in this article a microscopic derivation of the 
Bekenstein--Hawking entropy formula for four- and
five-dimensional non-extremal non-supersymmetric black holes. 
Previous successful attempts to count microscopic black-hole 
states were based on D-brane techniques, and were confined to 
extremal (or, at best, infinitesimally away from extremal) 
configurations, where part of supersymmetry is preserved (strictly
in the extremal limit).
For non-extremal black holes, the best attempts to date
only succeeded in deriving the correct dependence of the 
Bekenstein--Hawking entropy formula on the charges, but not the 
precise numerical coefficient. In this article we have computed the
entropy of 
non-extremal, non-supersymmetric $4d$ and $5d$ black holes 
from a microscopic point of view,
by embedding these black holes into M-theory and then
using its symmetries to map them, via a series of U-dualities, 
into configurations whose non-compact part 
is the three-dimensional BTZ black hole. We then performed a counting of 
microscopic states by following Carlip's approach. The latter is valid 
at and away from extremality. We furthermore argued that certain branes 
are dual to supersingleton representations of various gauged
supergravities. 
In this way we obtained a connection with the D-brane picture. 

A crucial step in our approach, that enabled us to relate solutions
of Poincar\'e and anti-de Sitter supergravities, was T-duality transformations
with respect to isometries which are space-like everywhere, except at spatial
infinity, where they become null. 
Moreover, since the 
non-compact time coordinate is involved in these transformations, the orbits
of the isometry are non-compact.
We have mentioned that a more 
proper treatment requires that we compactify the time (at some radius $R$) 
and at the end of the computation we send the compactification radius 
$R\to \infty$.
Notice that in the anti-de Sitter space that we obtain, after all dualities
have been performed, the time is naturally compact and taking the infinite
radius limit corresponds to considering the covering space.
Of course, then, the corresponding 
winding states decouple 
since they become infinitely heavy. However, it seems that 
these winding states should be projected out anyway since they are ghost-like,
i.e. the corresponding coordinate has the ``wrong'' sign in the action.
All non-dynamical processes (computation of the entropy is such a 
process) should be independent of whether or not one uses a non-compact time
or keeps the radius $R$
finite until the very end. Probably one should be more careful when it 
comes to fully dynamical processes, such as scattering. Nevertheless,
it will be interesting to understand this point better.

Higher-dimensional black holes also fall into our scheme, with one
exception: 
the eight-dimensional black holes. It will be
interesting to understand what distinguishes these black holes from the rest. 
The higher than five-dimensional 
black holes are associated to higher than three-dimensional
field theories; their analysis is therefore considerably harder.
In that respect, it would be interesting to better understand 
supersingleton field theories.
The latter have been analysed in \cite{NST}.
In that case the corresponding $p$-brane was considered to lie at 
the end of the world, where the topology is $S^1 \xx S^p$.
In the present context we would like to consider the boundary 
at the horizon of the black hole in a way similar to that used 
in section 3. Since the anti-de Sitter group coincides with 
the conformal group in one dimension lower, these field theories 
should be conformal field theories. The theory on the boundary 
of the BTZ black hole is indeed a conformal field theory.
Having obtained such field 
theories it would be desirable, as a next step, to further substantiate
our assessment that singletons account for the black-hole entropy of the 
black holes we have considered. For instance, for the case of
the $5d$ black hole, we may try to explicitly construct a
$(1+1)$ field theoretical action for them. This should arise 
from the synthesis of the six-dimensional and three-dimensional singleton 
actions corresponding to $adS_7$ and $adS_4$ spaces into a two-dimensional 
action corresponding to the one-dimensional intersection of the $M5$ and 
$M2$ branes. The resulting action should be related to (\ref{sbou}). Similar
considerations can also be made for the black holes corresponding to 
the spaces listed in Table 1, although for the higher than
five-dimensional black holes we do not know the form of the action in 
the intersection.

In this article we have only studied black holes that arise from
compactifications of type-II string theory. There are also heterotic black
holes. One might wonder whether our considerations apply in these as well.
Although we do not have a definite answer we remark 
that a mechanism that changes the asymptotics of $4d$ heterotic 
solutions (the corresponding 
non-extremal black-hole solutions have been constructed in \cite{cve})
has already been reported in \cite{bakas} for the 
gravity--dilaton--axion sector. Extensions of this work that will 
include the gauge fields should be important.

In our study of $4d$ and $5d$ black holes we have used the 
``BTZ gauge''. From Table 1 we see that there is also  
a U-dual configuration that involves, as the only non-compact part,
the two-dimensional $adS_2$ black hole. It has been shown \cite{bf} 
that the $adS_2$ gravity can be rewritten as a $BF$ theory, i.e.
a topological field theory. Therefore, one would expect 
that all degrees of freedom reside on the boundary, which is just a point.
Thus the computation of the entropy now becomes a quantum 
mechanical calculation.
What is truly remarkable is that, after the U-dualities, 
all the dependence of the entropy on the various charges resides
in the two-dimensional Newton constant. In this sense, the 
dependence of the Bekenstein--Hawking entropy formula on the mass 
and the charges is ``kinematical''.\footnote{This is 
in accordance with the fact that only qualitative considerations
are sufficient to determine
the correct dependence of the entropy on the various 
charges \cite{hor3, matrix}.}
The precise numerical coefficient
becomes a question that requires dynamics.
Such a calculation was performed in the last paper in \cite{carlipother}.
The authors reported negative results. It
view of the relevance of these results, it is definitely worth while to 
reexamine this calculation.

Perhaps the most interesting application is to study the 
final state of black holes in our framework.
In this respect the most promising ``gauge'' seems to be 
the $adS_2$ one. One should find U-duality-invariant 
quantities that uniquely characterize the final 
state of the black hole. Furthermore,
such calculations should involve the CFTs 
associated with the spheres and the tori since these
carry information about the original black hole.
We intend to return to this and other related issues in the future.


\bigskip\bigskip


\centerline{\bf Acknowledgements}

\noindent
Each of the authors wishes to thank the home institute of his co-author
for financial support and hospitality during crucial stages of this work.
K.Sk. is supported by the European Commission HCM program CHBG-CT94-0734 and
by European Commission TMR programme ERBFMRX-CT96-0045.

\appendix

\section{M-theory configurations}

In this appendix we present the 
M-theory configurations that yield, upon dimensional 
reduction in one coordinate, the $10d$ solutions 
discussed in section 2.

Consider the solution of the $11d$ supergravity that 
describes a non-extremal intersection of a five-brane ($M5$),
a membrane ($M2$) with a wave ($W$) in one of the common directions.  
Let us wrap  the $M5$ in  
$(x_1, x_2, x_3, x_4, x_5)$, the $M2$ in  
$(x_1, x_{10})$ and put a wave along the $x_1$ common direction.
The coordinates $x_i, i=1, \ldots, 5, 10$, are assumed to be periodic,
each with radius $R_i$. 
Explicitly, the solution is given by \cite{ts1}:
\bea
ds_{11}^2 &=& H_T^{1/3} H_F^{2/3} 
\Biggl(H_T^{-1} H_F^{-1} \left
(-K^{-1} f dt^2 + K \left(d x_1 + (K'{}^{-1} -1)dt\right )^2\right) \nonu
&&+H_F^{-1} (dx_2^2 + \cdots + dx_5^2) + f^{-1} dr^2 + r^2 d\O_3^2 
+ H_T^{-1} d x_{10}^2 \Biggl) ~ ,
\eea
with\footnote{The $*$-duality operation in (\ref{cf441}) and 
(\ref{cf442}) is defined with respect to (flat) transverse 
four-dimensional and three-dimensional spaces, respectively.}
\be
\cf_4 = -3 dt \wedge d H_T'{}^{-1} \wedge dx_1 \wedge dx_{10} + 
3 * d H_F' \wedge d x_{10} \label{cf441}~ ,
\ee
where the various harmonic functions are given by  
\bea \label{harm1}
&&K=1 + \frac{\cq_K}{r^2}, \  
K'{}^{-1} = 1 - \frac{Q_K}{r^2} K^{-1}, \ 
\cq_K=\m^2 \sinh^2 \a_K, \ 
Q_K= \m^2 \sinh \a_K \cosh \a_K \nonu
&&H_T=1 + \frac{\cq_T}{r^2}, \  
H'_T{}^{-1}= 1 - \frac{Q_T}{r^2} H_T^{-1}, \ 
\cq_T=\m^2 \sinh^2 \a_T, \ Q_T= \m^2 \sinh \a_T \cosh \a_T \nonu
&&H_F=1 + \frac{\cq_F}{r^2}, \  
H'_F= 1 + \frac{Q_F}{r^2}, \ 
\cq_F=\m^2 \sinh^2 \a_F, \ Q_F= \m^2 \sinh \a_F \cosh \a_F ~ ,
\eea
and $f$ is 
the same as in (\ref{fdef}) with $D=5$. The extreme limit is given by 
$\m \rightarrow 0, \a_K \to \infty, \a_T \to \infty, \a_F \to \infty$, while
$Q_K, Q_T$ and $Q_F$ are kept fixed. Upon dimensional reduction along $x_{10}$,
one obtains (\ref{10d}). In (\ref{10d}) we have renamed 
$H_T \to H_f$ and $H_F \to H_{s5}$, and also 
the charges,
so that it is clear to which brane each one of them is associated.

The M-theory configuration that yields (\ref{4d})
involves three five-branes wrapped in $(x_1, x_3, x_4, x_5, x_6)$, 
$(x_1, x_2, x_5, x_6, x_{10})$ and 
$(x_1, x_2, x_3, x_4, x_{10})$,
each intersecting at a threebrane, with a 
wave along the string common to all three branes in the direction $x_1$.
The metric and the four-form are given by \cite{ts1}
\bea
d s_{11}^2&=&(H_{F1} H_{F2} H_{F3})^{2/3}\biggl(
H^{-1}_{F1} H^{-1}_{F2} H^{-1}_{F3}
\left(-K^{-1} f dt^2 + K (d x_1 + (K'{}^{-1} -1)dt )^2\right ) \nonu
&&+H^{-1}_{F2} H^{-1}_{F3} (d x_2^2 + d x_{10}^2)
+H^{-1}_{F1} H^{-1}_{F3} (d x_3^2 + d x_4^2) 
+H^{-1}_{F1} H^{-1}_{F2} (d x_5^2 + d x_6^2) \nonu
&&+f^{-1} dr^2 + r^2 d \O_2^2 \biggr) ~ ,
\eea
and
\be
\cf_4=3(*d H'_{F1} \wedge d x_2 \wedge d x_{10} +
*d H'_{F2} \wedge d x_3 \wedge d x_4 +
*d H'_{F3} \wedge d x_5 \wedge d x_6)  \label{cf442} ~ .
\ee
The various harmonic functions are defined as 
\bea \label{harm2}
&&H_{Fi}=1+{\cq_{Fi} \over r} ,~~
H_{Fi}'=1+{Q_{Fi} \over r} ,~~   
\cq_{Fi}=\m\sinh^2\a_{Fi} , ~~ Q_{F_i} = \m\sinh\a_{Fi} \cosh\a_{Fi} ~ ,
\nonu 
&&K =1+{ \cq_K\over r} ,~~ 
K'{}^{-1}{=}1{-}{Q_K \over r}K ,~~
\cq_K=\m\sinh^2\a_K ,~~ Q_K=\m\sinh\a_K \cosh\a_K ~ ,
\eea
where $i=1, 2, 3$, and $f$ is the same as in (\ref{fdef}) with $D=4$. 
The extreme limit is given by $\m \rightarrow 0, 
\a_K \to \infty, \a_{Fi} \to \infty$, while $Q_K$ and $Q_{Fi}$, $i=1,2,3$,
are kept fixed.
Upon dimensional reduction along $x_{10}$ one obtains a solitonic five-brane
$NS5$ in $(x_1, x_3, x_4, x_5, x_6)$, 
two $D4$-branes in $(x_1, x_2, x_5, x_6)$ and $(x_1, x_2, x_3, x_4)$, 
with a wave along $x_1$. We further T-dualize this solution
along $T_{56}$. This yields the solution (\ref{4d}) used in section 2.
There, the harmonic functions are renamed as $H_{F1} \to H_{s5}$,
$H_{F2} \to H_{2}$ and $H_{F3} \to H_{6}$. The charges and the angles
are also appropriately renamed. 

\section{Higher-dimensional black holes\hfill\break
         and the BTZ black hole}

In this appendix we will show that the BTZ black hole can 
only be connected with $4d$ and $5d$ black holes and not with 
higher-dimensional ones.

Black holes arise from brane intersections after dimensional
reduction. The dimensionality $D$ of the final black hole 
is equal to the overall transverse dimension plus 1.
The solution depends on a number of harmonic functions
with respect to the overall transverse space, i.e. they are of the 
form
\be
H=1 + {Q \over r^{D-3}} ~ .
\ee
We consider non-extremal configurations obtained 
from extremal ones according to the rules discussed in \cite{ts1}.

As a first step we will rewrite the BTZ black hole metric (\ref{dsbtz})
in a way that depends on $(D-1)$-dimensional harmonic functions. 
To this end, consider the change of variables
\be
\r^2 = {4 \over (D-3)^2} {r^{D-3} \over \m^{D-5}} + \r_-^2 ~ .
\ee
We also make the following identifications
\bea \label{ident}
&& l = {2 \over (D-3)} \m~ ,~~~~
\r_+ = l \cosh \a~ ,~~~~
\r_- = l \sinh \a~ , 
\nonumber \\
&& \varphi = {x \over l}~ ,~~~~
H = {\m^{D-3} \over r^{D-3}}~ .
\eea
The final result is that the BTZ metric (\ref{dsbtz}) takes the form
\bea
 \label{bbtz1}
d s^2_{BTZ} & = & H^{-1} \left(-K^{-1}(r) f(r) dt^2 + K(r) \left(dx + 
(K'{}^{-1}(r)-1)dt\right)^2 \right) \nonumber \\
&& + H^{{2 \over D-3}} f^{-1}(r) dr^2 ~ ,
\eea
where $f$, $K$ and $K'$ are as in (\ref{fdef}).

Let us now examine whether or not this metric can result 
from an intersection of branes. If this is the case, the form of $H$ implies
that the BTZ 
black hole will emerge after the shift transformation 
has been applied. 
The configurations we examine are built 
from superpositions of single brane solutions. The latter 
have the form
\be
ds^2 = H^{\a_p} \left( H^{-1} (-dt^2 + dx_1^2 + \cdots dx_p^2) 
+ (d x_{p+1}^2 + \cdots + dx_{D-1}^2) \right) ~ ,
\ee
where $H$ is a harmonic function and $a_p$ is a numerical factor that depends 
on the particular brane (e.g. $\a_2 = 1/3$ for the $M2$, $\a_5 = 2/3$
for the $M5$, etc.). What is important for our discussion is
that the difference between the power of the harmonic 
function multiplying the overall transverse coordinates and 
the power of the harmonic function multiplying the 
world-volume coordinates is 1. This is true
for all branes (and also for the wave). 
If one superimposes two branes, then the 
difference of the sums of the powers will be 2.
For a solution that depends on $k$ charges (the charges may 
be degenerate) the difference will be equal to $k$.
The same results hold for non-extremal intersections built according to the 
rules of \cite{ts1}. Notice also that this number is preserved
in the process of dimensional reduction. 
The explicit form of a solution arising from the intersection of $k$ branes is 
\be
ds^2 = \left(\prod_{i=1}^{k} H_i^{a_i}\right) 
\left(\left[-\left(\prod_{i=1}^{k} H_i\inv\right) f dt^2 + d s^2_{w}\right] 
+ d s^2_{RT}
+ f\inv dr^2 + r^2 d\O^2_{D-2}  \right)~ ,
\label{gjh}
\ee
where $d s^2_{w}$ denotes the 
metric of the world-volume coordinates apart from the $dt^2$-term shown in
(\ref{gjh}),
and $d s^2_{RT}$ is the metric of the relative transverse coordinates.
The precise form of the latter is irrelevant for our argument.
Here 
$H_i$ are harmonic functions (including the one associated with the wave) and
$a_i$'s are some numerical constants.  
Therefore, if the solution (\ref{bbtz1}) originates from an 
intersection of branes, then we should have
\be
{2 \over D-3} + 2 = k~ .
\ee
This equation has solutions (for integers $D$ and $k$) only for
$D=4, k=4$ and $D=5, k=3$. These are precisely the cases studied in 
section 2, namely for a $4d$ black hole that depends on four charges and
a $5d$ black hole that depends on three charges. Notice that this 
argument depends crucially on the fact that we build our 
intersection according to the rules of \cite{ts1}, namely  
we start from supersymmetric configurations and then add 
deformation terms appropriately. For a supersymmetric
configuration, $k$ is always an integer \cite{kinteg}.


\end{document}